\documentclass[a4paper]{jpconf}
\usepackage{graphicx}
\usepackage[square,sort&compress, numbers]{natbib}
\newcommand{\txs}{TXS~0506+056}

\begin{document}
\title{Neutrinos from blazars}

\author{Matteo Cerruti}

\address{Institut de Ci\`{e}ncies del Cosmos (ICCUB), Universitat de Barcelona (IEEC-UB),\\ Carrer de Mart\'{i} i Franqu\`{e}s 1, E08028 Barcelona, Spain}

\ead{matteo.cerruti@icc.ub.edu}

\begin{abstract}
The evidence for joint gamma and neutrino emission from the blazar \txs\ has renewed interest in blazars as neutrino sources. The detection of neutrinos from blazars can be seen as the smoking gun for the presence of relativistic protons in blazar's jets, and can thus help identify blazars, and active galactic nuclei in general, as cosmic-ray accelerators. In this contribution I first introduce blazars and blazar hadronic models, and then present the results of the multi-messenger modeling of both the 2017 gamma-neutrino flare, and the 2014-2015 neutrino-only flare of \txs. For the first time it is possible to constrain blazar hadronic models using the information from neutrino detectors, and I discuss the implications of these neutrino observations on our understanding of the physics of relativistic jets from super-massive black-holes. 
\end{abstract}

\section{Blazars}

The identification of the sites of cosmic-ray acceleration in the Universe 
is among the most important open questions in astrophysics \citep{Blandford14, Blasi13}. Among the different candidates, active galactic nuclei (AGNs) have long been considered of particular interest. AGNs are the observational effect of the accretion of matter onto super-massive black holes (SMBHs, with $M_\bullet \simeq 10^{8-9} M_\odot$) that dwell in the center of galaxies \citep{EHT}. The accretion of matter takes the form of an accretion disk, that is surrounded at a larger scale by a dusty torus. Clouds of matter that orbit the SMBH are ionized by the disk black-body radiation, and emit recombination lines that can be observed in the optical/UV spectrum of AGNs. The lines are more or less Doppler-broadened as a function of the distance from the SMBH, and we talk about a broad-line region (BLR) and a narrow-line region (NLR). The large variety of AGN observational properties are interpreted as being due to the inclination of the system with respect to the observer, with the central region around the SMBH visible if the system is seen face-on, or obscured by the torus if the system is seen edge-on. In a sub-class (around 10$\%$) of AGNs, the accretion onto the black hole is associated with an outflow in the form of a pair of relativistic jets of plasma that are launched along the polar axis of the system. Synchrotron emission from the relativistic jets makes these AGNs bright in the radio band, hence their name \textit{radio-loud} AGNs. When the relativistic jet points towards the observer, relativistic effects boost the emission and makes these objects among the brightest sources of photons in the Universe. Within the AGN unified scenario \citep{Urry95}, we call \textit{blazars} these peculiar radio-loud AGNs. \\

From an observational point of view, blazars are characterized by the presence of a spectral non-thermal continuum in optical/UV, a high degree of polarization, super-luminal motion in radio, and rapid variability (down to time-scales of minutes) at all wavelengths. All these properties are well understood within the scenario described above, in which the emission is associated with a non-thermal population of particles in a jet of plasma that is moving relativistically towards the observer. Blazars are not composed of a homogeneous population, and they are further divided into two subclasses that carry the historical names of \textit{BL Lacertae objects} (BL Lacs) and \textit{Flat-Spectrum Radio-Quasars} (FSRQs). The classification is done on the basis of the optical/UV spectrum of the source: if it is characterized by a non-thermal continuum, without emission/absorption lines, the source belongs to the subclass of BL Lacs, while, if emission lines from the BLR are observed, the source is classified as an FSRQ. The two sub-classes indicate the presence of two intrinsically different inflow/outflow regimes, which impact the ratio of non-thermal to thermal emission. This dichotomy matches the dichotomy also observed in the parent population of radio-galaxies \citep{Fanaroff74}, with powerful jets seen in FR II (associated with FSRQs) and fainter jets seen in FR I (associated with BL Lacs).\\

The spectral energy distribution (SED, that is the representation of the energy density $\nu F_\nu$ as a function of $\nu$) of blazars is characterized by a non-thermal continuum from radio to very-high-energy $\gamma$-rays (VHE, $E>100$ GeV), composed of two distinct components: the first one peaks in infrared-to-X-rays, while the second one peaks in the $\gamma$-ray band, from MeV-to-TeV energies. FSRQs are all characterized by a low-frequency of the SED peaks, which are located in the infrared and the MeV band, respectively. On the other hand, BL Lacs show a variety of peak frequencies and can be further classified into low/intermediate/high-frequency peaked BL Lacs (LBLs/IBLs/HBLs), if the frequency of the first SED peak is below 10$^{14}$ Hz, between 10$^{14}$ and 10$^{15}$ Hz, and above 10$^{15}$ Hz, respectively \citep{Padovani95}. 
The physics driving the properties of blazar SEDs is not clear : a first study by \citep{Fossati98} identified an anti-correlation between the luminosity and the frequency of the first SED peak, the so-called \textit{blazar sequence}. According to it, the brightest blazars (the FSRQs) have the lowest peak frequencies, and moving to lower luminosities the peak frequencies increase, up to HBLs. Several works have investigated whether the blazar sequence is intrinsic to blazar physics, or due to observational biases. Several outliers to the blazar sequence have been identified.  More recent studies have evolved the original sequence into a more complex blazar \textit{envelope} \citep{Meyer11}. Recently the blazar sequence has been revisited by \citep{Ghisellini17}.\\
The origin of the low-energy SED component in blazars is unanimously ascribed to synchrotron emission by a population of leptons (electrons/positrons) in the jet. This result is solid, thanks mainly to the detection of polarized emission in blazars. The spectral shape of the emission is also consistent with what is theoretically expected from synchrotron emission, assuming that the underlying particle distribution is in the form of a power-law (or broken-power-law). The origin of the high-energy SED component is more disputed. In leptonic models, it is associated with inverse-Compton scattering off low-energy photons by the same electrons/positrons that are responsible for the low-energy SED component. If the low-energy photons are the synchrotron photons by the same leptons, the emission is called Synchrotron-Self-Compton (SSC); if on the other hand the low-energy photons are external to the relativistic jet, we talk about External-Inverse-Compton (EIC). Typical external fields that are used in EIC modeling are the thermal photons from the SMBH accretion disk, the emission lines from the BLR, or the thermal photons from the dusty torus. Given that this low energy photon fields are directly observed in FSRQs, the EIC scenario is usually applied (successfully) to the modeling of FSRQs, while the SSC scenario works better for HBLs.

\section{Blazar hadronic models and neutrinos from AGN}
As an alternative to leptonic models, hadronic models associate the high-energy SED component with a non-thermal population of hadrons (protons) in the jet. Protons can radiate directly via synchrotron radiation, or indirectly via secondary particles produced in proton-photon interactions. As in leptonic models, the soft photons that act as a target can be internal or external to the emitting region. The first p-$\gamma$ process to be considered in blazar hadronic models is the photo-meson channel, that is the production of pions, both neutral and charged. Neutral pions produced in the interaction decay directly into photons, while charged pions decay into leptons (muons and then electrons/positrons) and neutrinos. The neutrinos produced in this process are a unique signature for the presence of p-$\gamma$ interactions in the jet, and thus for the acceleration of hadrons (and thus cosmic-rays) in the relativistic jets of AGNs.  Photons and secondary leptons from pion decay triggers synchrotron-supported pair-cascades in the jet: electrons/positrons pairs are produced in $\gamma$-$\gamma$ interactions, these secondary pairs radiate synchrotron photons which pair-produce again, and so on. This cascade emission transfers part of the radiative power to lower energies, and emerges as a secondary radiative component in the SED. Another process which is relevant in p-$\gamma$ interactions is the Bethe-Heitler pair production, that is the direct production of an electron/positron pair. This Bethe-Heitler pairs, similarly to the leptons from the pion decay, trigger a pair cascade in the emitting region. Finally, in some part of the parameter space, synchrotron emission by muons, produced in the decay of charged pions, and before their decay into electrons/positrons, can emerge as an additional high-energy radiative component. Neutrinos can also be produced in proton-proton interactions. This process is usually negligible in AGN jets, due to the low density of the plasma, but can become dominant if the target protons are from an obstacle encountered by the jet, such as BLR clouds, or stars \citep{Barkov12}. \\

\begin{figure}[t!]
\begin{center}
\includegraphics[width=30pc]{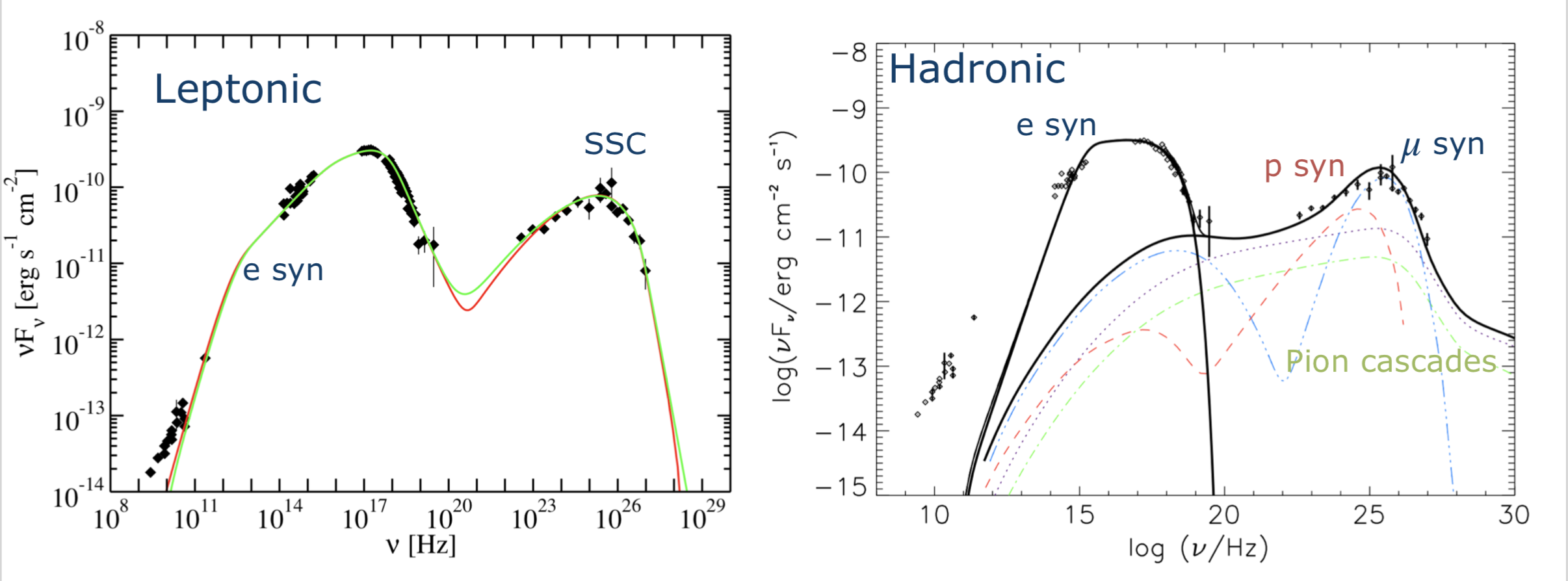}
\caption{\label{figone}Spectral energy distribution of the HBL Markarian 421, modeled in a leptonic (left) and hadronic (right) scenario. Adapted from \citep{mrk421fermi}}
\end{center}
\end{figure}

Blazar hadronic models have been developed well before the rise of neutrino astronomy \citep{Mannheim93, Aharonian00, Mucke01, Cerruti15, Petropoulou15}: besides their interest for the natural link they provide with cosmic-ray physics, it is important to remind that blazar leptonic models do not always provide a good description of blazar SEDs. The best example of such "problematic" blazar observations are the so-called \textit{orphan} flares, observed only in one energy band without counterparts at other energies. These events are very difficult to explain if the main radiative process is inverse Compton scattering (due to the direct connection between the two SED components in this scenario). Hadronic models provide in general a good description of blazar SEDs, as good as the leptonic ones, and it is very difficult to discriminate between the two scenarios \citep{Bottcher13}. One difficulty encountered by hadronic models is that the power associated with a hadronic solution is much larger than the one needed by a leptonic solution, due to the larger mass of the proton. In some cases the power exceeds the Eddington luminosity ($L_{\rm{Edd}}$) of the SMBH. Although $L_{\rm{Edd}}$ should be seen only as a rough estimation of the available accretion power (this is particularly true for blazar for which we do not have an estimate of $M_\bullet$), a power in the jet much larger (orders of magnitudes) than $L_{\rm{Edd}}$ raises doubts on the feasibility of these radiative models \citep{Zdziarski15}. This problem is particularly relevant for hadronic modeling of FSRQs, while for HBLs is it generally possible to find hadronic solutions that do not face energetic issues. An example of leptonic versus hadronic blazar modeling is shown in Fig. \ref{figone} for the HBL Markarian 421 \citep{mrk421fermi}. 
 
\section{TXS~0506+056}
\txs\ is a $\gamma$-ray blazar which was relatively unknown until September 22, 2017. On that day, the IceCube neutrino detector observed a high-energy ($E \simeq 290$ TeV) neutrino,  IC-170922A, coming from a direction consistent with this blazar \citep{TXS0506}. A multi-wavelength campaign revealed that \txs\ was ongoing a 6-month long $\gamma$-ray flare, as observed with Fermi-LAT, and VHE $\gamma$-rays were also detected with the MAGIC Cherenkov telescopes. The chance coincidence of the two events to occur simultaneously has been estimated at the $3$-$\sigma$ level. Prompted by this 2017 event, the IceCube collaborations searched for neutrinos consistent with the position of \txs, and identified a $3.5$-$\sigma$-significant neutrino flare during 2014-2015 \citep{TXS05062014}. Interestingly enough, this "historic" neutrino flare does not seem to have been accompanied by a flare in the electromagnetic band. Although none of the events is, by itself, above the $5$-$\sigma$ golden standard, they are the most significant evidence, as per today, of joint photon-neutrino emission from a blazar, and thus for the acceleration of cosmic rays in AGN jets. Since 2017, \txs\ has attracted a major interest in the AGN community. First, and most importantly, the redshift of the source (unknown at the time of the IceCube alert) has been estimated as $z=0.3365$ \citep{Paiano18}. The source has then been studied in details at all wavelengths, and an interesting result is that \txs\ does not seem to follow the standard blazar sequence described above, and stands out as an atypical over-luminous IBL/HBL \citep{Padovani19}. \txs\ is naturally the best candidate to test blazar hadronic models: for the first time it has become possible to fit not just the electromagnetic SED, but also the neutrino SED, putting new constraints on blazar radiative models. In the following I will discuss what did we learn from the (evidence of) neutrino emission from \txs, both from the 2014 and 2017 events, and what is the impact of these observations on blazar models.

\section{The 2017 gamma-neutrino event from TXS~0506+056}
The September 2017 event has triggered a considerable effort in the blazar community to try to explain the photon and neutrino emission from \txs. In order of complexity, the first models investigated are the so-called one-zone scenarios, in which the whole emission is ascribed to a single emitting region in the relativistic jet. In the simplest one-zone model, the low-energy photons that act as a target for p-$\gamma$ interactions are internal to the emitting region, and primarily are synchrotron photons from the leptons that produce the low-energy SED component. \\

In the proton-synchrotron scenario the bulk of the high-energy SED component is directly associated with proton synchrotron photons, with negligible contributions from other radiative components. Several authors have shown, independently, that the 2017 flare from \txs\ cannot be described by a proton-synchrotron model: although the electromagnetic SED can be well fitted, the expected neutrino rate is too low to be consistent with the IceCube detection\footnote{This remains true even when considering the role of the Eddington bias when comparing the model neutrino rates to the IceCube detection, see \citep{edbias}} \citep{Cerruti19, Keivani18, Gao19}. This is due to the fact that typical proton synchrotron solutions are characterized by a high magnetic field (to let the proton synchrotron component dominate) and a low-density emitting region (to suppress the SSC component). They imply a low rate of photo-meson production, and thus of neutrinos. This is a very important conclusion from a single neutrino detection, and shows that a whole family of solutions that was considered viable until 2017, cannot fit the new multi-messenger informations we have (at least for this particular source).\\

Alternatively to pure hadronic solutions, mixed lepto-hadronic solutions have been proposed. In this case, the bulk of the high-energy component is ascribed to SSC photons, with a sub-dominant hadronic component that emerges (as Bethe-Heitler and pion cascades) in X-rays and VHE $\gamma$-rays. In this case the emitting region is much denser (also to allow an efficient SSC production) and the expected neutrino rates observed with IceCube are of the order of $0.1$ neutrinos per year, consistent with the neutrino detection. The drawback of these solutions is that the required power is of the order of $10^{48-50}$ erg s$^{-1}$, much larger than the Eddington luminosity of the SMBH that powers the system (which is around $10^{46-47}$ erg s$^{-1}$ for a black hole with $M_\bullet \simeq 10^{8-9} M_\odot$).
The energetic requirement can be lowered if the p-$\gamma$ interactions happen not on photons within the emitting region, but on external ones. This can be easily understood considering that, if there are more photons with the right energy, and thus the probability for a p-$\gamma$ interaction increases, we need less protons to get the same photon and neutrino output. This kind of lepto-hadronic solutions, characterized by an electromagnetic SED dominated by SSC radiation, with a sub-dominant hadronic component produced over an external photon field, are capable of explaining both the photon and neutrino emission of \txs, with a reasonable energy budget \citep{Keivani18, Gao19, Ansoldi18, Righi19}. The natural question that arises is then: what is the external field over which the neutrinos are produced? \txs\ is a BL Lac, so the BLR/torus photons that are normally considered for FSRQs are not an easy answer. However, being an unusually luminous BL Lac, it may be possible that an external field bright enough is still present in the AGN environment. Alternatively, the external field may come from a structured jet, in which the $\gamma$-$\nu$ emitting region is embedded in a larger sheath that produces the target photons.
With a joint photon and neutrino modeling, we can now constrain for the first time the proton population in the relativistic jet of an AGN. Again, independently, different research groups converged on the fact that \txs\ (and thus by extension AGNs during similar flares) does not accelerate cosmic rays up to ultra-high-energies. The maximum proton energy, which is a free parameter for the various models, is constrained to be lower than $10^{18-19}$ eV, and thus cannot reach the highest energies of the cosmic ray spectrum observed at Earth. 
The jet/obstacle radiative model has also been applied to the 2017 flare of \txs, showing that this kind of scenario is capable of fitting the electromagnetic SED and at the same time producing a neutrino flux in line with the IceCube detection \citep{Wang18, Liu19, Sahak18}.

\begin{figure}[t!]
\begin{center}
\includegraphics[width=14pc]{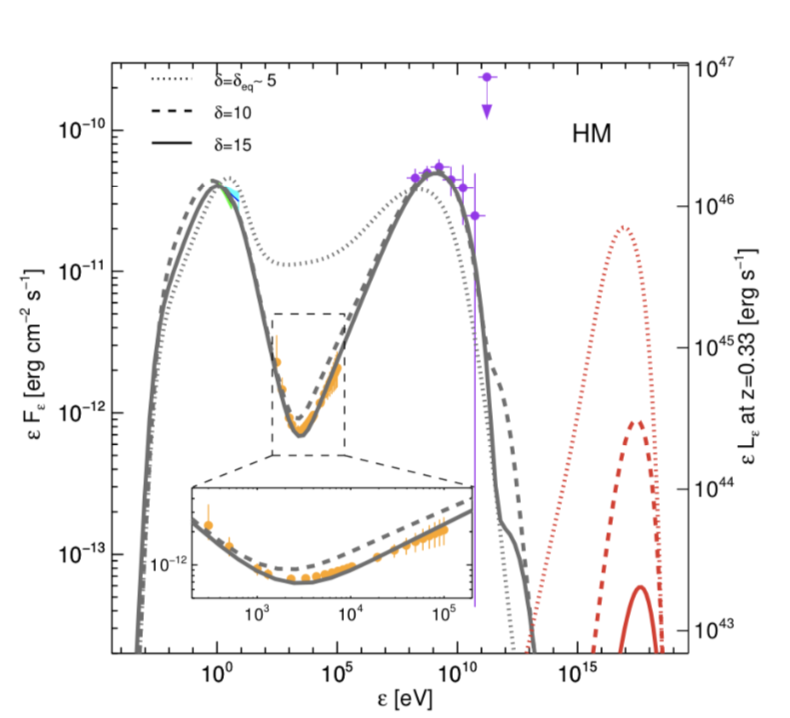}
\includegraphics[width=14pc]{figtwob.png}\\
\includegraphics[width=14pc]{figtwoc.png}
\includegraphics[width=14pc]{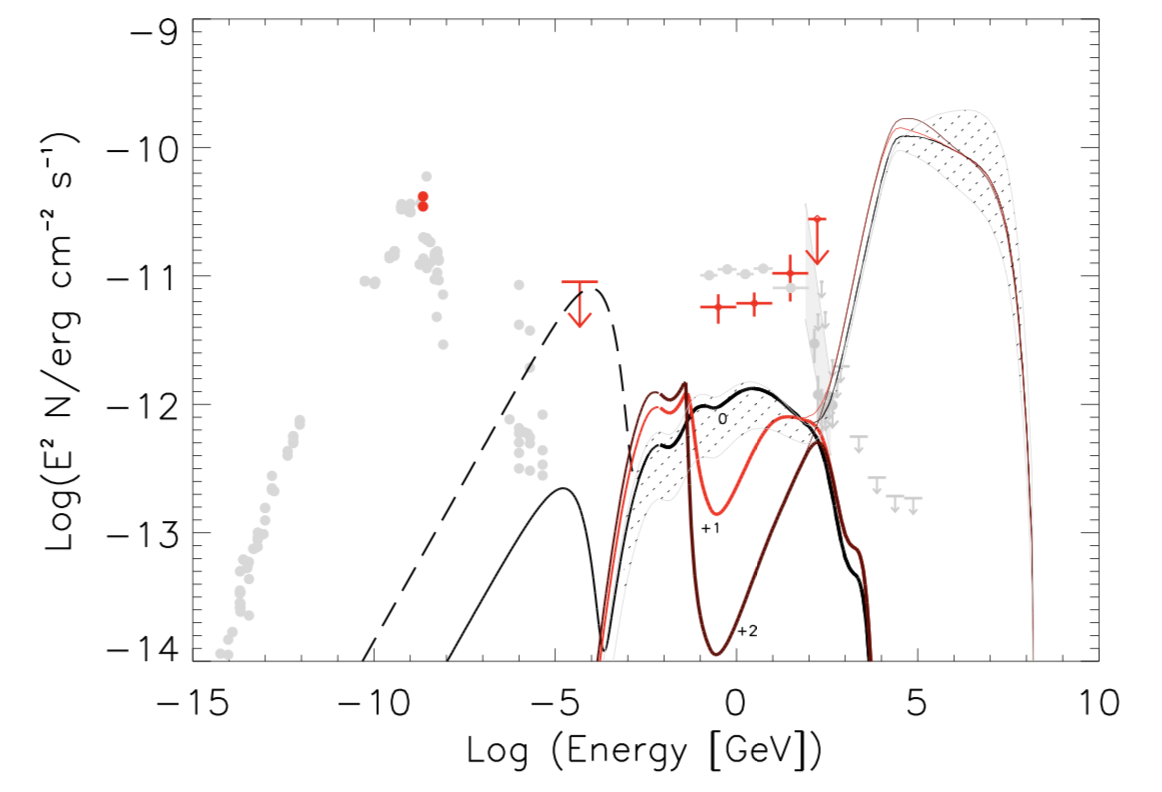}\\
\caption{\label{figtwo} Top: spectral energy distribution of the \txs\ during the 2017 event modeled in a proton synchrotron (left, from \citep{Keivani18}) and lepto-hadronic (right, from \citep{Gao19}) scenario. Bottom: spectral energy distribution of the \txs\ during the 2014 event modeled in a single-zone (left,from \citep{Rodrigues19}) and inverse-Compton-supported cascade (right, from \citep{Reimer19}) scenario. }
\end{center}
\end{figure}

\section{The 2014-2015 neutrino flare from TXS~0506+056}
An evidence ($3.5$-$\sigma$) for a  neutrino flare from \txs\ during 2014-2015 has been reported by IceCube. In this case the electromagnetic coverage is much worse than for the 2017 event, due to the absence of a specific alert at that time, and the best we can do is to work with survey instruments such as Fermi-LAT. With the information we have, we can say that during the neutrino flare there was no photon flare going on. This event is, if possible, even more interesting than the 2017 one. As discussed above, in one-zone blazar hadronic models the photon and neutrino emission is intimately related due to their co-production in p-$\gamma$ interactions. If neutrinos are produced, then photons have to be produced as well. In case the emitting region is opaque to $\gamma$-$\gamma$ pair-production, the energy of the photons can be redistributed via pair-cascades, but it cannot disappear. As discussed in \citep{Rodrigues19}, the 2014 event is extremely constraining for one-zone models: if we want to fit the neutrino emission without violating electromagnetic observations we need to store the cascade emission into the only energy band which is currently poorly covered and has the worst upper limit, the MeV band. Although this option cannot be excluded, it seems unlikely, mainly because the resulting SED has never been observed in any blazar. \\

The evolution of the pair cascade under different hypotheses  is investigated in details in \citep{Reimer19}. The authors concluded that while synchrotron supported cascades clearly violate the electromagnetic constraints, it is possible to have inverse-Compton-supported cascades. In this case however, the neutrino and the photon emission has to be associated with different emitting regions in the relativistic jet, breaking the gamma-neutrino connection. A similar way out is discussed by \citep{Murase18} in the \textit{neutron-beam} scenario: neutrons produced in p-$\gamma$ interactions escape without energy losses from the photon emitting region, and produce neutrinos further away. 

\section{Perspectives}
Even though multi-messenger observations have shown (evidence of) joint gamma-neutrino emission from \txs, it does not mean that all blazars emit neutrino, nor that blazars are at the origin of the diffuse neutrino background observed by IceCube. As discussed in \citep{ICLAT}, $\gamma$-ray blazars, that dominate the LAT diffuse background, cannot produce more than 10-30$\%$ of the neutrino background. On the other hand, the 2014 \txs\ neutrino flare may suggest the existence of neutrino emission from blazars without an electromagnetic counterpart. These orphan neutrino events may contribute significantly to the neutrino background \citep{Halzen19}.\\

An important question we need to answer is why the first neutrino AGN has been \txs. As discussed earlier, it is an unusually luminous blazar that stands out within the blazar population. In addition, the six-months long flare associated with the 2017 high-energy neutrino is unusual among $\gamma$-ray blazars, and can also have played a role. We should, however, try not to be biased by our current knowledge of blazars. It may be possible (as hinted by the 2014 event) that there exist neutrino (hadronic) blazar flares and photon (leptonic) flares, and in general there may be hadronic-dominated blazars and leptonic-dominated blazars. The population of neutrino blazars might thus do not overlap with the population of photon blazars. Neutrino observations can thus give us new and independent informations on the physics of AGN jets, and on the properties of the hadronic accelerators at work in them. \\
  
The multi-messenger observations of \txs\ gave us some important observational lessons: the first one, from the 2017 event, is that the hadronic emission associated with the IceCube neutrino emerges in the hard-X-ray and the VHE $\gamma$-ray band, highlighting the importance of these two energy bands in multi-wavelength campaigns; the second one, from the 2014 event, is that there is a huge need for an MeV satellite that can monitor the transient Universe similarly to Fermi-LAT at higher energies. Future constraints will come certainly from the detection (or not) of additional blazar neutrinos with IceCube or Antares, but also from the continued monitoring of \txs\ at all wavelengths \citep{VERITAS0506, KonstICRC}.  

\ack{
M. Cerruti has received financial support through the Postdoctoral Junior Leader Fellowship Programme from la Caixa Banking Foundation, grant n. LCF/BQ/LI18/11630012. Funding for this work was partially provided by the Spanish MINECO under project MDM-2014-0369 of ICCUB (Unidad de Excelencia 'Mar\'{i}a de Maeztu')
}

\bibliography{Cerruti_TAUP}

\end{document}